\documentclass[reprint,aps,epsfig,superscriptaddress,twocolumn]{revtex4-1}
\usepackage{bm}
\usepackage[dvips]{graphicx}
\usepackage{mathrsfs}
\usepackage[intlimits]{amsmath}
\usepackage{textcomp}
\usepackage[switch]{lineno}
\usepackage[colorlinks, citecolor=red]{hyperref}
\usepackage{tabularx}

\begin{document}

\title{500$\,$Hz atom-atom entanglement distribution over metropolitan-scale fibers}

\author{J.-X. Shi}
\altaffiliation{These authors contributed equally to this work}
\affiliation{Center for Quantum Information, Institute for Interdisciplinary Information Sciences, Tsinghua University, Beijing 100084, PR China}

\author{Y.-D. Sun}
\altaffiliation{These authors contributed equally to this work}
\affiliation{Center for Quantum Information, Institute for Interdisciplinary Information Sciences, Tsinghua University, Beijing 100084, PR China}

\author{Y.-B. Liang}
\affiliation{Center for Quantum Information, Institute for Interdisciplinary Information Sciences, Tsinghua University, Beijing 100084, PR China}

\author{S. Zhang}
\affiliation{State Key Laboratory for Artificial Microstructure and Mesoscopic Physics and Frontiers Science Center for Nano-optoelectronics, School of Physics, Peking University, Beijing, 100871, China}

\author{P.-Y. Liu}
\affiliation{Center for Quantum Information, Institute for Interdisciplinary Information Sciences, Tsinghua University, Beijing 100084, PR China}

\author{Z.-B. Cui}
\affiliation{Center for Quantum Information, Institute for Interdisciplinary Information Sciences, Tsinghua University, Beijing 100084, PR China}

\author{L.-M. Duan}
\email{lmduan@tsinghua.edu.cn}
\affiliation{Center for Quantum Information, Institute for Interdisciplinary Information Sciences, Tsinghua University, Beijing 100084, PR China}
\affiliation{Hefei National Laboratory, Hefei 230088, PR China}

\author{Y.-F. Pu}
\email{puyf@tsinghua.edu.cn}
\affiliation{Center for Quantum Information, Institute for Interdisciplinary Information Sciences, Tsinghua University, Beijing 100084, PR China}
\affiliation{Hefei National Laboratory, Hefei 230088, PR China}


\begin{abstract}
Quantum networks and repeaters are among the most promising routes to scale up quantum information systems. Nevertheless, various aspects still require significant improvement to achieve practical realization. Multiplexing can enhance the performance in many fundamental steps, thus emerging as a critical strategy for the realization of quantum repeaters. Currently, a major bottleneck in realizing functional quantum repeaters is heralded entanglement distribution over long-distance elementary links, as neither of the two dominant protocols---single-photon nor two-photon interference---can simultaneously achieve high efficiency and strong robustness. Here, we address this challenge by experimentally realizing a method that unifies the efficiency and robustness. By harnessing two $15\times15$ quantum memory arrays, we generate remote matter-matter entanglement at a rate of $546\,$Hz over $20\,$km of fiber without active phase stabilization. This rate exceeds that of two-photon and single-photon schemes by factors of $75$,$932$ and $23$, respectively, establishing a new benchmark for entangling speed over metropolitan-scale distances and enabling semi-deterministic entanglement distribution. Inherently compatible with multiplexed quantum repeater architectures, our scheme offers a promising approach towards the realization of future high-performance distributed quantum systems.
\end{abstract}

\maketitle

\section*{Introduction}

Quantum repeaters~\cite{BDCZ, DLCZ, duan_arxiv, rmp_gisin, duan_RMP} serve as the communication backbone of a future global quantum internet~\cite{kimble_internet,hanson_internet}, opening up applications such as distributed quantum computing~\cite{distributed,duan_pra,oxford_distributed,rempe}, networked quantum sensing~\cite{repeater_telescope,oxford_clock,lukin_repeater_telescope}, and long-distance quantum communication~\cite{weinfurter_qkd, bao_qkd,wangye_qkd}. Although remarkable progress on quantum repeaters has been reported recently~\cite{230m,hanson_long,weinfurter_qkd, bao_qkd,wangye_qkd, 2021nature, zhou,ion-ion}, significant efforts are still required to realize a functional quantum repeater capable of distributing entanglement over a continental distance. For instance, current single-mode quantum repeaters achieve remote entanglement generation rates of approximately $1\,$Hz  over $10\,$km of fibers~\cite{wangye_qkd,bao_qkd}. Given that direct secure key distribution over hundreds of kilometers has already achieved rates of tens of hertz~\cite{zhoulai}, such low-rate links fail to deliver the anticipated acceleration of quantum repeaters.

To enhance the performance of a quantum repeater, multiplexing schemes have been proposed and demonstrated~\cite{2007prl, rmp_gisin, 2021nature, zhou, tittel_frequency, Oblak, oam,lan,225,multipurpose,enhancement,changwei,wavevector,lanyon_multiplexing,haffner_multiplexing, solid_ion, cui, ion-ion, fiber_array}. A multiplexed quantum repeater with a high-capacity on-demand quantum memory in each node can significantly boost the performance in many key aspects. As illustrated in Fig.~1, this architecture not only accelerates entanglement generation over elementary links~\cite{ion-ion,2021nature,zhou,cui,enhancement} but also enables the storage and synchronization of multiple remote entangled pairs, which is essential for operations such as entanglement swapping~\cite{lanyon_connection,lattice} and purification~\cite{hanson_purification,riedmatten_purification,DLCZ}. Therefore, multi-mode operation is essential for a quantum repeater. Currently, the crucial step for demonstrating  scalability---synchronization of remote matter-matter entangled pairs heralded at random times---remains unrealized yet at metropolitan scales.

Another primary requirement for practical quantum repeaters is the efficient and high-fidelity generation of heralded entanglement over an elementary link.  Typically, there are two schemes for heralded entanglement generation: single-photon interference (SPI)~\cite{zoller_SPI,DLCZ,bao_qkd,wangye_qkd,blatt_SPI} and two-photon interference (TPI)~\cite{ion-ion,oxford_distributed, 230m}, both of which are the current workhorses in experimental realizations. The SPI scheme offers high efficiency, as the heralding of successful entanglement relies on the detection of only a single photon, as illustrated in Fig.~1. Nevertheless, this approach is constrained by a drawback: the need for rigorous phase stabilization, which requires actively locking the optical path length to within a fraction of a wavelength. In contrast, TPI has a much more relaxed phase stability requirement (approximately $10^7$ less stringent), as one only needs to match the temporal envelopes of the two photons. Nevertheless, TPI suffers from low efficiency, since two photons must be registered simultaneously for a successful heralding event. Therefore, both schemes possess intrinsic limitations that constrain the performance of entanglement generation in an elementary link, the basic step for a quantum repeater.

\begin{figure*}
    \centering
    \includegraphics[width=\textwidth]{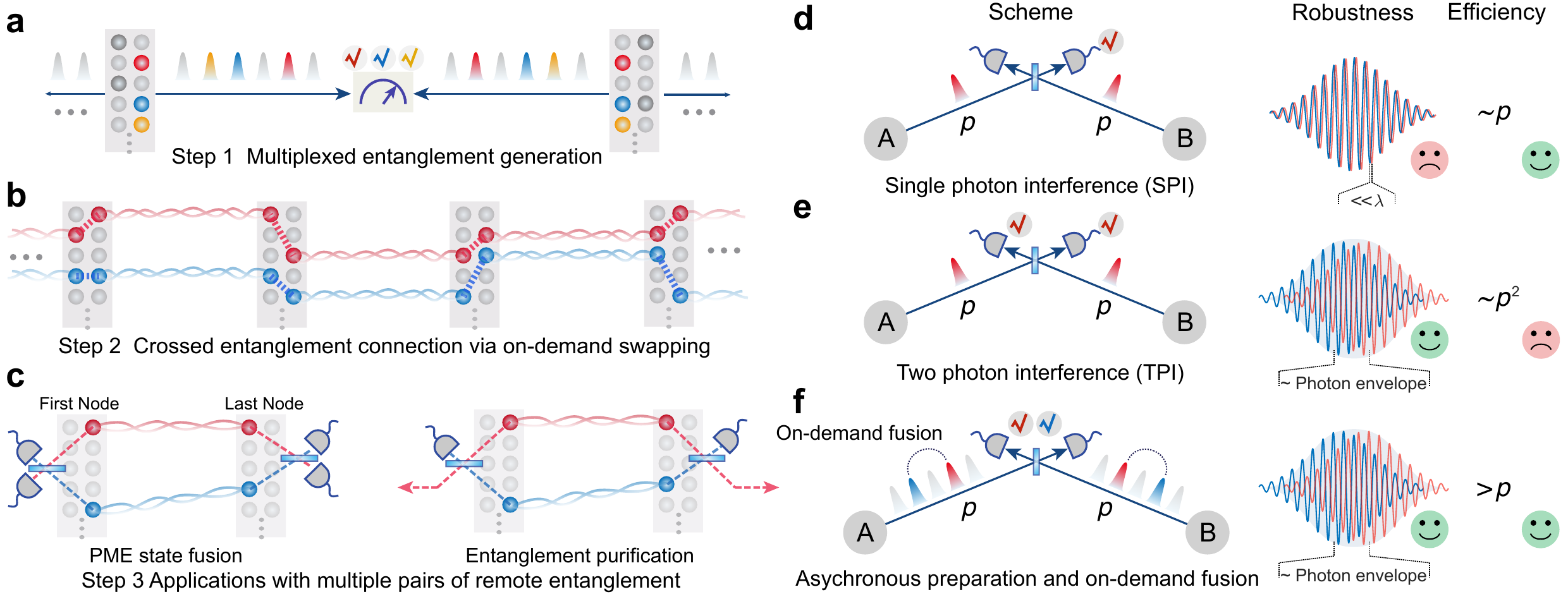}
    \caption{
    \textbf{Multiplexed quantum repeater and schemes for entanglement generation.}
     \textbf{a}, The rate of entanglement generation over an elementary repeater link is substantially enhanced through concurrent entangling attempts, which circumvent the inherent round-trip communication latency. In addition, the multiplexed quantum repeater architecture enables the generation and storage of multiple remote entangled pairs.
     \textbf{b}, To extend the spatial extent of quantum entanglement, entanglement swapping can be performed on two cells within the same quantum memory array, connecting asynchronously generated remote entanglements into a larger one. In a multiplexed quantum repeater consisting of $M$ nodes and an $N$-cell on-demand quantum memory array in each node, the multiplexing scheme improves the entanglement distribution rate by $N^{M-1}$ relative to a single-mode repeater~\cite{2007prl}.
     \textbf{c}, PME states can be generated by fusing two EME states following the DLCZ protocol in a multiplexed quantum repeater. Entanglement purification~\cite{riedmatten_purification} and other applications requiring multiple pairs of remote entanglement can be readily implemented within a multiplexed quantum repeater architecture.
     \textbf{d}, SPI is efficient but requires stringent optical phase stabilization. Active phase stabilization with high precision ($\Delta L\ll\lambda$) is essential, which poses significant challenges when the optical path length reaches tens of kilometers.
     \textbf{e}, TPI is inefficient but robust in phase. It merely requires the overlap of the two photon envelopes, a condition that is far less stringent than that of SPI.
     \textbf{f}, A scheme simultaneously incorporates the high efficiency of SPI and the high robustness of TPI by asynchronously generating two EME states via SPI and fusing them into a PME state via on-demand quantum memory.
    }
    \label{Fig1}
\end{figure*}

In this work, we experimentally realize an entanglement generation scheme that combines the high efficiency of SPI with the high robustness of TPI, based on two random-access quantum memory arrays~\cite{225,multipurpose} within a multiplexed quantum repeater architecture. Our scheme starts by asynchronously generating two effective maximally entangled (EME) states between remote memory arrays via SPI, using a series of successive excitations on the memory cells at both nodes. Driven by multiplexing enhancement, we achieve an EME  state generation rate of $1.6\,$kHz across $20\,$km of telecom fiber. We then use real-time feedforward of heralding signals and on-demand memory retrieval to synchronize and fuse these EME states into a polarization maximally entangled (PME) state, following the DLCZ (Duan-Lukin-Cirac-Zoller) protocol~\cite{DLCZ}. Enabled by the ultra-fast generation of the two EME states via multiplexing, the optical phase remains unchanged during their creation. This ensures a stable phase in the resulting PME state, eliminating the challenging requirement for active phase stabilization. Utilizing on-demand quantum memory to harness all the randomly generated EME pairs, we achieve a remote PME state generation rate of $546\,$Hz over $20\,$km of fiber. This is $75$,$932$ ($23$) times faster than the single-mode TPI (SPI) schemes, establishing a new record for metropolitan-scale entanglement generation rate. Additionally, the entanglement generation success probability reaches $30\%$ in the heralding stage, approaching the performance of deterministic remote entanglement generation. Therefore, by combining multiplexing and memory enhancements, the fast and robust entangling scheme demonstrated in this work provides an important building block for realizing practical quantum repeaters.

\begin{figure*}
	\includegraphics[width=0.85\linewidth]{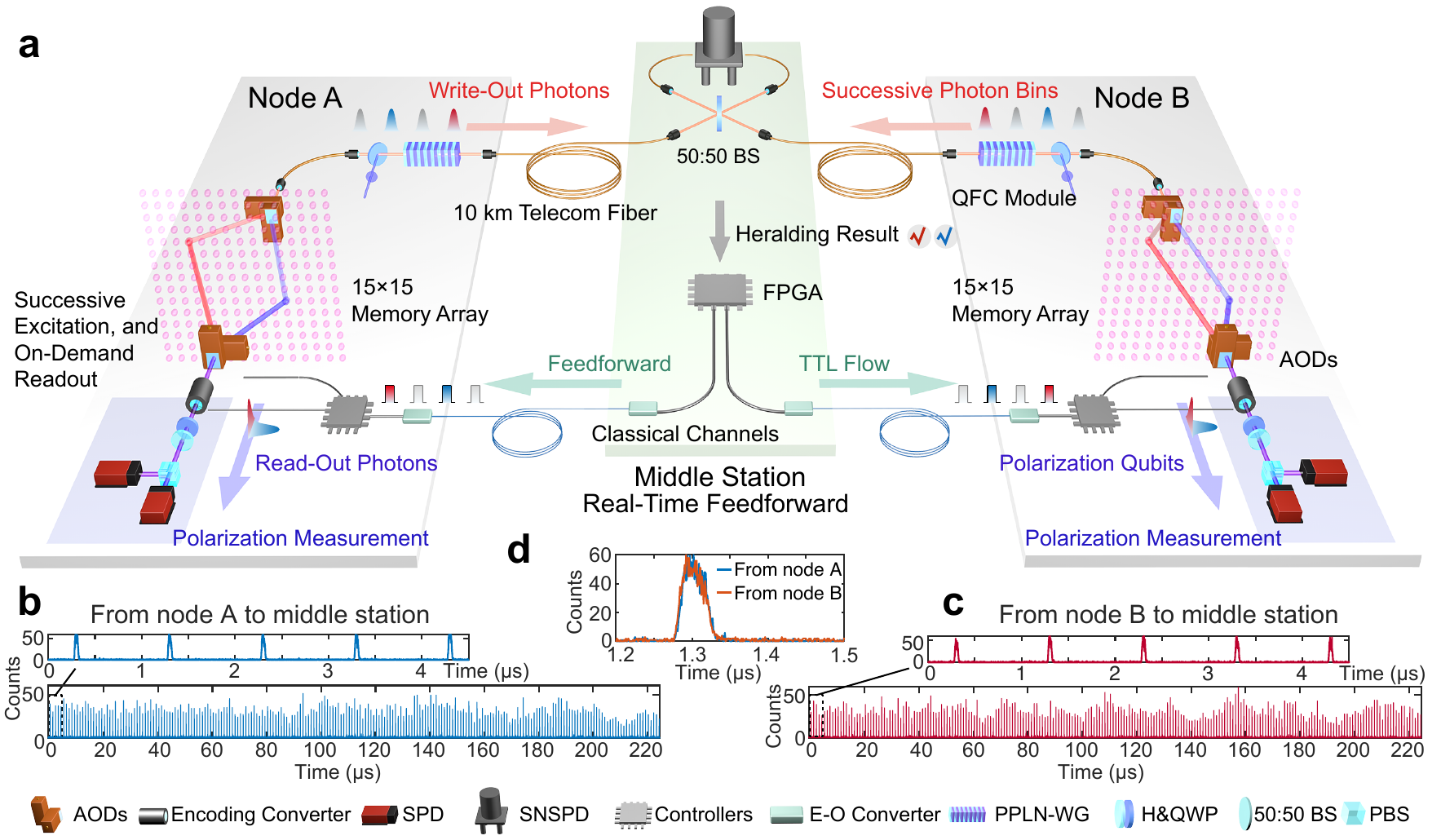}
	\caption{\textbf{Experimental setup and multiplexed entanglement generation.}
	\textbf{a}, The experimental setup consists of two quantum repeater nodes, labeled A and B, and a middle measurement station. Both nodes are linked to the middle station through $10\,$km spools of telecom fiber. Each node is equipped with a $15 \times 15$ random-access quantum memory array, operating at a clock rate of $1\,$MHz~\cite{multipurpose}. In each excitation attempt, the emitted write-out photon is converted to the telecom band within a quantum frequency conversion module and subsequently sent through the long fiber to the middle station. Photons from nodes A and B interfere at a $50$:$50$ beamsplitter and are detected by two superconducting nanowire single-photon detectors (SNSPDs) to herald the EME state between remote quantum memory cells via SPI. The heralding signals (TTLs) are transmitted back to nodes A and B via electro-optic (E-O) converters and additional long fibers. The returned TTL signal heralds the successful generation of a memory-memory EME state, and the control units identify the specific pair of entangled memory cells based on the arrival time of the heralding signals.
	\textbf{b-c}, One excitation round consists of successive excitations of $225$ cells within the memory arrays in each node, spaced by $1\,\mu$s. The histogram of photon arrival events is measured at the middle station.
	\textbf{d}, The write-out photons from both nodes have a pulse width of $50\,$ns and exhibit good temporal overlap at the beamsplitter.
	}
	\label{Fig2}
\end{figure*}

\section*{Experimental protocol and results}

We build a two-node quantum repeater with a long-lived high-capacity quantum memory array in each node. As shown in Fig.~2, the quantum memory is based on a two-dimensional ($15\times15$) array of $^{87}$Rb micro-ensembles, with each micro-ensemble serving as an independent quantum memory cell. Each cell can be individually addressed by a scanning system based on four pairs of acoustic optical deflectors (AODs), enabling unconstrained random-access operations~\cite{multipurpose}. With the co-propagating configuration of the write beam and write-out photon, the coherence times of the two quantum memory arrays (averaged over $225$ cells) are $683\,\mu$s and $503\,\mu$s for node A and B, respectively. Thanks to the random-access capability, stored quantum information can be read out at any desired time within the memory lifetime, fulfilling the essential on-demand retrieval requirement for quantum repeaters.

The entanglement generation between two distant quantum memory arrays starts with successive excitations of the $225$ memory cells in each node. We execute this by one or two rounds for heralding. In each excitation, a $50\,$ns write pulse drives spontaneous Raman scattering with a small probability, yielding a forward-scattered $795\,$nm write-out photon and a spin-wave excitation, following the DLCZ protocol~\cite{DLCZ}. The excitation of each memory cell takes $1\,\mu$s, resulting in a total duration of $225\,\mu$s for one complete excitation round. The write-out photons are first converted into $1542\,$nm via DFG (difference frequency generation) process in a PPLN (Periodically-poled Lithium Niobate) waveguide. Then the converted write-out photons in the telecom band are sent through a $10\,$km optical fiber to the middle measurement station. The histogram of the write-out photon modes for node A and B is shown in Fig.~2.

\begin{figure*}
\centering
\includegraphics[width=\linewidth]{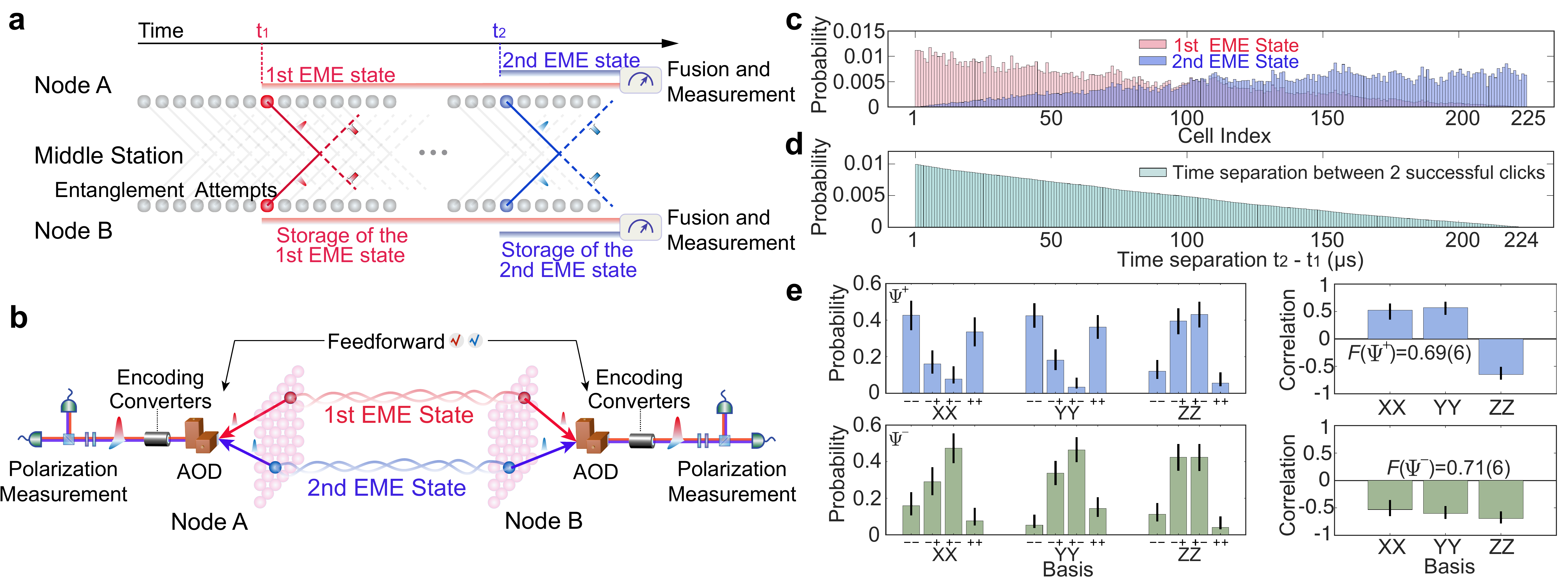}
\caption{\textbf{Experimental protocol and results.}
  \textbf{a}, Asynchronous generation of two EME states via SPI in the multiplexed excitations.
  \textbf{b}, Fusing two asynchronously generated EME states into a PME state. The control unit calculates which cells are excited according to the arrival time of the heralding signals. Then the two cells excited at $t_1$ and $t_2$ are retrieved out as read-out photons one after another with a time separation of $1\,\mu$s. An encoding converter based on an actively-locked asymmetric Mach-Zehnder interferometer and an Electro-optical modulator further converts the read-out photons from time-bin encoding to polarization encoding. The yielded PME state is further measured via waveplates and PBSs.
  \textbf{c}, Distribution of cell index regarding the first and second EME states heralded at random times $t_1$ and $t_2$ in one round of excitations (See Supplementary material for the detailed cell indexes).
  \textbf{d}, Distribution of the time separation $t_2-t_1$ for the two asynchronous EME states.
  \textbf{e}, Entanglement fidelities for the PME states. The measurement results for the correlations XX, YY, ZZ and the fidelities with Bell states $|\Psi^{+}\rangle$ and $|\Psi^{-}\rangle$ are demonstrated for $|\Psi^{+}\rangle_{\text{PME}}$ and $|\Psi^{-}\rangle_{\text{PME}}$, respectively.
}
\label{Fig3}
\end{figure*}

Once the write-out photons from nodes A and B arrive at the middle measurement station, they are interfered on a $50$:$50$ beamsplitter. A single photon click in either of the two detectors heralds the successful generation of an EME entangled state between nodes A and B, encoded in the Fock basis as $|\Psi\rangle_{\text{EME}}=(|1\rangle_A|0\rangle_B\pm e^{i\phi}|0\rangle_A|1\rangle_B)$. Here $|0\rangle_{A/B}$ and $|1\rangle_{A/B}$ denote the absence or presence of an atomic spin-wave excitation in the corresponding memory cell, respectively. The sign $\pm$ is determined by the specific heralding detector, while $\phi$ is an unknown relative phase determined by the lasers and the overall optical paths, including the long fibers.

In the heralding stage, we asynchronously generate two EME entangled states $|\Psi_{t_1}\rangle_{\text{EME}}=(|1\rangle_A|0\rangle_B\pm e^{i\phi_1}|0\rangle_A|1\rangle_B)/\sqrt{2}$ and $|\Psi_{t_2}\rangle_{\text{EME}}=(|1\rangle_A|0\rangle_B\pm e^{i\phi_2}|0\rangle_A|1\rangle_B)/\sqrt{2}$ between A and B at random time $t_1$ and $t_2$ via SPI. The distribution of $t_1$, $t_2$, and $t_2-t_1$ are demonstrated in Fig.~3.

These two asynchronously generated EME states are further synchronized and fused into a PME state via feedforward~\cite{DLCZ,kimble_science}, based on the heralding signal timing and the on-demand storage of the first EME state until the second occurs, as shown in Fig.~3. Thanks to the on-demand quantum memory array with arbitrarily programmable storage times, all $\binom{N}{2}\propto N^2$ ($N$ is the number of total excitations) EME pairs can be harnessed to create a memory-memory PME state, thereby achieving a full acceleration in entangling rate. The yielded memory-memory PME state can be expressed as
\begin{equation}
|\Psi^{\pm}\rangle_{\text{PME}}=\frac{1}{\sqrt{2}}(|t_1\rangle_A|t_2\rangle_B\pm e^{i\Delta\phi}|t_2\rangle_A|t_1\rangle_B)
\end{equation}
where $|t_1\rangle$ and $|t_2\rangle$ denote the memory cell excited at time $t_1$ and $t_2$ in each node, and $\Delta\phi=\phi_{1}-\phi_{2}$ is the phase of the PME state after fusion. Since the time interval $t_2-t_1$ is significantly smaller than the characteristic time of optical path fluctuation in our experiment (see the analysis later in this paper), we have $\phi_{1}\approx\phi_{2}$, and thus $\Delta\phi\approx0$. This mechanism ensures phase stability without the need for active stabilization of the long optical paths.

\begin{figure}
\centering
\includegraphics[width=0.48\textwidth]{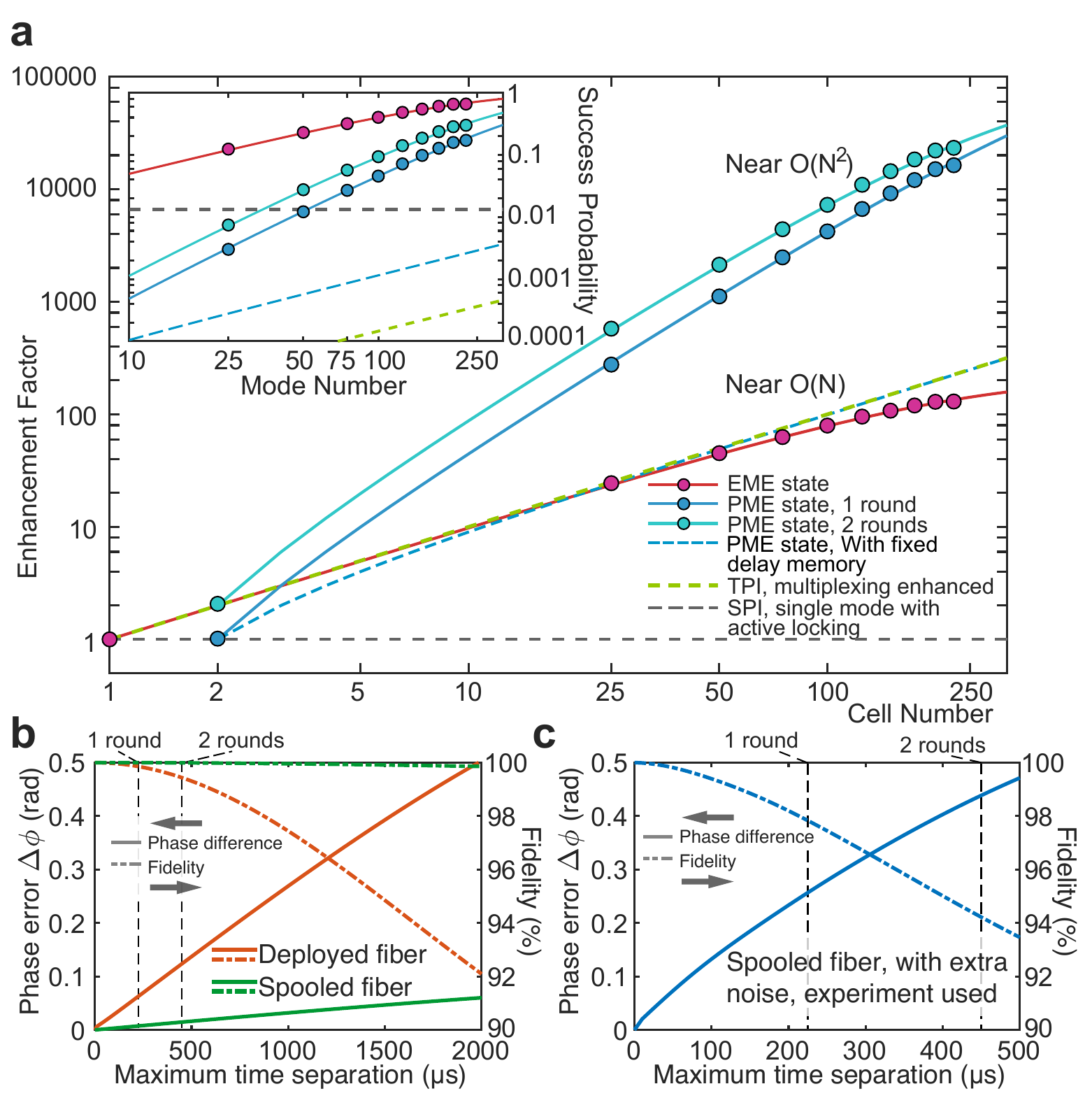}
  \caption{\textbf{Efficiency and robustness analysis.}
  \textbf{a}, We compare the multiplexing enhancement factor and remote entangling efficiency (inset) across different schemes as a function of the number of memory cells. The purple points and curve represent the measured data points and simulation of generating the intermediate EME state in two rounds of excitations via the scheme in our current experiment. The dark (light) blue points and curve represent the data points and simulation of generating the PME state in one (two) excitation rounds via the scheme in our current experiment (scheme (i)). The gray dashed curve represents the performance of generating remote entanglements via single-mode SPI with active phase locking (scheme (ii)). The green dashed curve represents the simulation of generating remote entanglement via multiplexing-enhanced TPI (scheme (iii)).  The blue dashed curve represent the simulation of generating PME states via the scheme (iv) (with only fixed-delay quantum memory). We apply corresponding adjustments based on the duration of each scheme to make them comparable.
  \textbf{b}, Measured standard deviation of phase uncertainty $\Delta \phi$ and inferred infidelity as a function of the maximum diffusion time, for both $22\,$km deployed and $20\,$km spooled fibers.
  \textbf{c}, The phase uncertainty and infidelity for the case with extra vibration noise.
  }
\label{Fig4}
\end{figure}

The yielded memory-memory PME state is converted to a photonic maximally entangled state by shining two successive read pulses to the corresponding memory cells $|t_1\rangle$ and $|t_2\rangle$. The time-bin encoding of the read-out photon is further converted to polarization encoding by a converter (Fig.~3b). After the conversion, the entangled state can be expressed as
\begin{equation}
|\Psi_{\text{read-out}}^{\pm}\rangle=\frac{1}{\sqrt{2}}(|H\rangle_A|V\rangle_B\pm |V\rangle_A|H\rangle_B)
\end{equation}
where $|H\rangle_{A/B}$ and $|V\rangle_{A/B}$ denotes H and V polarizations of read-out photons in node A or B.

The remote entanglement is further characterized using measurement bases implemented with waveplates and PBSs. Here we obtain the fidelity of the heralded memory-memory entanglement with respect to the maximally entangled Bell states $|\Psi^{\pm}\rangle$ by measuring the correlations $\langle \sigma_i^A\sigma_i^B \rangle, i \in \{x,y,z\}$, which are simplified as XX, YY, and ZZ. The measured correlations with one round of excitations and a success probability of $0.241\%$ per excitation are shown in Fig.~3, yielding a memory-memory entanglement fidelity of $\mathcal{F}=\frac{1}{4}(1+\text{XX}+\text{YY}-\text{ZZ})=69(6)\%$ for $|\Psi^{+}\rangle$ and $\mathcal{F}=\frac{1}{4}(1-\text{XX}-\text{YY}-\text{ZZ})=71(6)\%$ for $|\Psi^{-}\rangle$, respectively.

\section*{Performance analysis}

The acceleration of our method originates from  multiplexing and memory enhancements. Theoretically, our current scheme has a success probability of $\binom{N}{2}p^2(1-p)^{N-2}\propto N^2p^2$, where $p$ is the heralding probability of single-mode SPI. As in the current experiment $N^2p\gg 1$, our scheme is already faster than the single-mode SPI scheme, which exhibits an efficiency proportional to $p$. If $Np>1$ is further achieved ($Np$ approximates the success probability of generating a single EME state in the heralding stage), our scheme transitions into the deterministic regime, in which the remote entanglement can be heralded with a high probability. In our current experiment, $Np=1.08$ is achieved by running two rounds of excitations with an SPI success probability of $0.241\%$ per excitation. In this case, a success probability of generating a PME state as high as $30.02\%$ is measured, illustrating that our schemes is running in a semi-deterministic style.

Here we compare the efficiency of four different schemes for remote entanglement generation: (i) on-demand fusion of two asynchronous EME states (scheme employed in this experiment), (ii) single-mode SPI with global phase stabilization~\cite{bao_qkd,wangye_qkd}, (iii) multiplexed TPI~\cite{zhou, ion-ion}, and (iv) fusion of two EME states with fixed delay (similar to scheme (i) but with pre-programmed fixed-delay quantum memory)~\cite{zhou2026}. We plot the multiplexing enhancement factor and the probability of successful heralding for the four schemes in Fig.~4. Here we vary the number of memory cells used for the multiplexed schemes (i, iii, and iv), with only one cell used for the single-mode scheme (ii). Other conditions are the same with the current experiment. If all the cells are used, we are able to achieve a generation rate of $1.6\,$kHz for the EME state and $546\,$Hz for the PME state. The entangling efficiency of our current scheme is approximately two orders of magnitude higher than that of schemes (iii and iv), $23$ times higher than scheme (ii), and $7.59\times10^4$ times higher than single-mode TPI scheme. (The rates for different protocols are rescaled by duration for fair comparison.)

Here on-demand quantum memory array is essential for the efficient generation of remote entanglement. With on-demand quantum memory allowing for arbitrary storage times, any pair of asynchronous EME states $(t_1,t_2)$ can be utilized to yield a success probability $\propto N^2p^2$. In contrast, if a quantum memory with a fixed storage time is used, only pairs with a fixed $t_2-t_1$ can be utilized, resulting in a degraded success probability $\propto Np^2$ and significantly reduced efficiency. Furthermore, the lack of independent, on-demand addressing capabilities prevents scheme (iv) from connecting and purifying remote entangled pairs heralded at different times, rendering it difficult to sclae~\cite{2007prl}.

Regarding robustness, the time difference between the two EME states $t_2-t_1$ must remain within the characteristic time of fiber fluctuation, as the optical phase noise is dominated by the long fibers in our experiment. Here we characterize the fiber noise by measuring the optical phase uncertainty $\Delta\phi$ as a function of the maximum diffusion time, in three different cases. In the first two cases---specifically, $22\,$km deployed fibers buried under the Tsinghua University campus and $20\,$km of spooled fibers placed in a lab---the fiber-induced infidelities are negligible (Fig.~4b). To test the robustness of our scheme, we deliberately apply strong vibrations to the $20\,$km spooled fibers in the third case (Fig.~4c). The fiber noise in this scenario is significantly stronger than that in the first two cases and reaches a level comparable to the noise induced by a $250.2\,$km deployed fiber~\cite{riedmatten_fiber}. In our experiment, we intentionally select this noisiest channel (the third case) for the generation of the PME state (results shown in Fig.~3e and 3f), to demonstrate that our scheme remains operational even in an environment significantly noisier than realistic scenarios. While the induced infidelity reaches $2\%$ in this most challenging scenario (still considered acceptable), it remains below $0.5\%$ for the $22\,$km deployed fiber under the realistic conditions.

\section*{Conclusion and outlook}

In this work, we have realized an efficient and robust method for distributing heralded entanglement between distant quantum repeater nodes. Through this approach, we successfully generate heralded entanglement between two quantum memory arrays at a fast rate of $546\,$Hz over $20\,$km of noisy optical channels, establishing a new rate record for metropolitan-scale quantum repeaters. Notably, our experimental entangling efficiency significantly surpasses that of the conventional single-photon interference scheme and approaches the regime of deterministic remote entanglement generation. By unifying multiplexing and memory enhancements, we also demonstrate the synchronization of remote matter-matter entangled pairs heralded at random times---a critical capability for a scalable quantum repeater. Furthermore, active phase stabilization is not needed in this scheme, which significantly reduces experimental overhead and complexity. In addition, our experiment also provides a potential solution to the phase stability challenges in future multiplexed quantum repeaters based on tweezed atom arrays~\cite{fiber_array}, to avoid the difficulty of simultaneously phase-locking hundreds of optical paths. In the future, improvements in the quantum memory and addressing units via optical lattice arrays, faster AOD switching, and integration of other dimensions~\cite{wavevector,oam} are expected to enable deterministic entanglement distribution over hundreds of kilometers without active phase locking. This approach can also be extended to deployed fibers~\cite{bernien,linke,230m,zhou2026,weinfurter_qkd}, facilitating entanglement distribution between spatially separated locations. The implementation of this efficient and robust method for generating distant quantum entanglement thus could become an important building block for future large-scale quantum networks.\\

\noindent
\textbf{Acknowledgements:}
We acknowledge helpful discussions with Yong Yu.
\textbf{Funding:} This work is supported by Innovation Program for Quantum Science and
Technology (No.2021ZD0301102), the Tsinghua University
Initiative Scientific Research Program and the Ministry of
Education of China through its fund to the IIIS. Y.F.P. acknowledges
support from the Dushi Program
from Tsinghua University.
\textbf{Author Contributions:}  J.X.S., Y.D.S., Y.B.L., S. Z.
and Y.F.P. carried out the experiment. L.M.D. and Y.F.P. supervised
the project. All the authors contributed to the discussion and
the writing.
\textbf{Competing interests:} The authors declare that
there are no competing interests.
\textbf{Data and materials availability:}
All data related to this paper will be released at the time of publication.

\onecolumngrid

\setcounter{MaxMatrixCols}{12}

\setcounter{equation}{0}
\setcounter{figure}{0}
\setcounter{table}{0}
\setcounter{page}{1}
\setcounter{section}{0}
\makeatletter
\renewcommand{\thefigure}{\textbf{S\arabic{figure}}}
\renewcommand{\figurename}{\textbf{Fig.}}
\renewcommand{\theequation}{S\arabic{equation}}
\renewcommand{\thetable}{S\arabic{table}}

\pagebreak
\begin{center}
\Large Supplemental Material for\\\textbf{500$\,$Hz atom-atom entanglement distribution over metropolitan-scale fibers}
\end{center}

\subsection*{Section S1. Random-access Quantum Memory Array}

In this section, we detail the implementation and performance of the quantum memory array at each node. Within the memory array, the atomic ensemble is cooled to approximately $30\,\mu\text{K}$ using the $780\,\text{nm}$ transitions via an optical molasses process. It is then divided into spatially separated sub-ensembles arranged in a $15 \times 15$ array within the same plane. Each memory cell (sub-ensemble) has a diameter of approximately $100\,\mu\text{m}$, with an array spacing of $160\,\mu\text{m}$. Every cell can be independently addressed for on-demand read and write access, facilitating the implementation of the Duan-Lukin-Cirac-Zoller (DLCZ) quantum repeater scheme. The two nodes are located in the same lab, separated by $5\,$m spatially.

The configuration of the quantum memory array is illustrated in Fig.~S1a. We address different cells in the memory array using four sets of crossed acousto-optic deflectors (AODs), which correspond to the write beam, the write-out signal photon, the read beam, and the read-out idler photon. The write and read control beams are combined with the signal and idler single-photon paths using a beamsplitter with $95\%$ transmittance. This achieves a strictly collinear geometry, ensuring that the spin-wave memory suffers minimal decoherence induced by atomic thermal motion. The arrangement of the quantum memory array requires an AOD addressing frequency spacing of $2.5\ \text{MHz}$ within a bandwidth ranging from $85\ \text{MHz}$ to $120\ \text{MHz}$, as illustrated in Fig. S1b. During continuous operation, switching the AOD optical path to address a specific mode takes $800\ \text{ns}$, resulting in a total operation time of $1\,\mu\text{s}$ per unit.

The key performance metrics of the memory arrays at both nodes are presented in Fig.~S1c-e. The retrieval efficiencies at Nodes A and B are shown in Fig.~S1c and S1d, respectively, with individual memory unit efficiencies generally ranging from $10\%$ to $20\%$. The average storage efficiency is $17.9\%$ for Node A and $13.9\%$ for Node B. These efficiencies account for the transmittance of the optical components in the array (such as the AODs) but exclude the efficiencies of the subsequent channel components and detectors. Variations in storage efficiency are primarily attributed to slight differences in the optical depth of each memory unit.
\begin{figure}
	\includegraphics[width=1\linewidth]{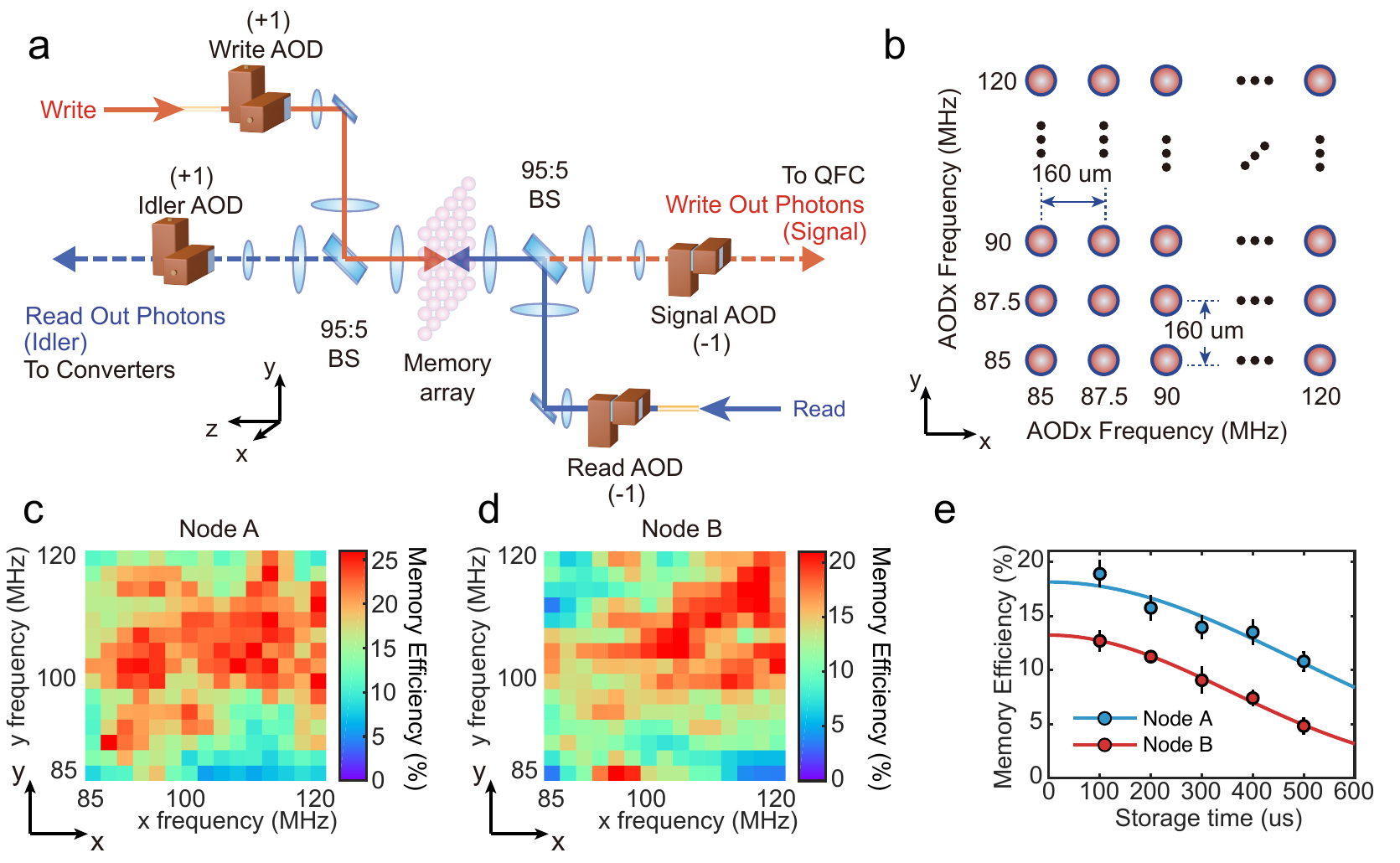}
	\caption{\textbf{Experimental setup of the quantum memory array.}
		\textbf{a}, Optical layout of the memory array at each node.
		\textbf{b}, Spatial distribution of the memory array and the corresponding AOD addressing frequencies.
		\textbf{c}, Storage efficiency of the memory array at Node A.
		\textbf{d}, Storage efficiency of the memory array at Node B.
		\textbf{e}, Storage efficiency decay of the memory array over time and the coherence time fit.
	}
	\label{FigS1}
\end{figure}
\begin{figure}
	\includegraphics[width=\linewidth]{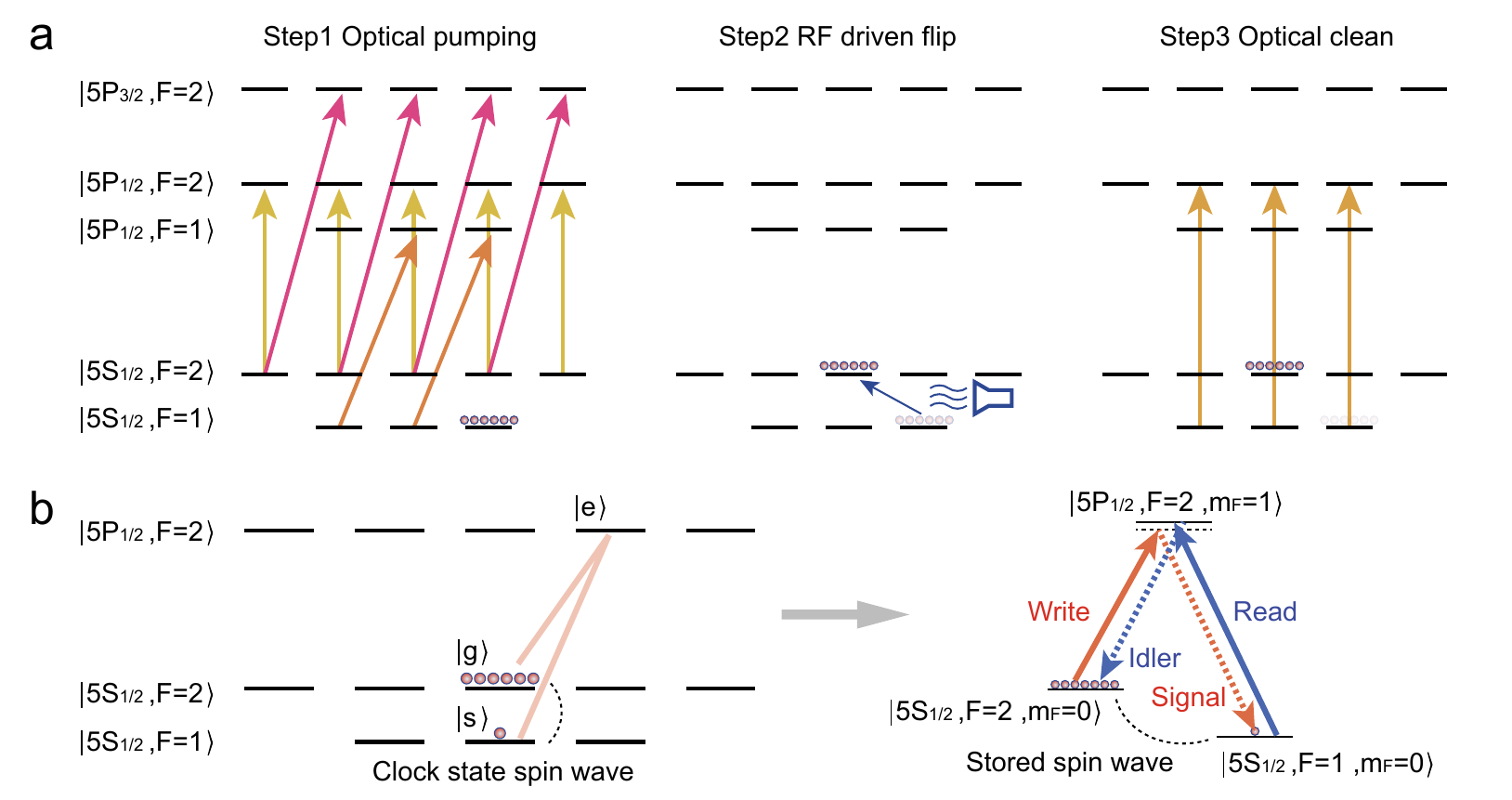}
	\caption{\textbf{Energy levels used in the DLCZ scheme.}
		\textbf{a}, Scheme for initial state preparation.
		\textbf{b}, Energy level structure of the DLCZ scheme.
	}
	\label{FigS2}
\end{figure}
Coherence time measurements for the memory arrays are shown in Fig.~S1e. By executing a dynamic read-write sequence, we measured the retrieval efficiency of the array at various storage times. Fitting the data with the function $\eta(t) = \eta \exp(-(t/\tau)^2)$ yields a coherence time of $\tau_A = 683\,\mu\text{s}$ for Node A and $\tau_B = 503\,\mu\text{s}$ for Node B, resulting in an average storage time of $593\,\mu\text{s}$ across the two nodes. Compared to previous works \cite{225,multipurpose,enhancement}, the memory array realized in this experiment achieves over $200$ independently accessible units while maintaining a long coherence time exceeding $500\,\mu\text{s}$. This duration is sufficient to complete two full rounds of multiplexed excitations plus the $100\,\mu\text{s}$ round-trip transmission time through $20\,\text{km}$ of optical fibers.

The energy level structure utilized in the quantum repeater scheme is depicted in Fig.~S2. The initial state of the atomic ensemble, $|5S_{1/2}, F = 2, m_F = 0\rangle$, is prepared via a three-step process (Fig.~2a). The first step involves a $10\ \mu\text{s}$ optical pumping process. Under a $0.5\ \text{G}$ magnetic field, we simultaneously activate three pumping beams (one $\sigma^+$ beam coupling the $|5S_{1/2}, F = 1\rangle \rightarrow |5P_{1/2}, F = 1\rangle$ transition, one $\pi$ beam coupling the $|5S_{1/2}, F = 2\rangle \rightarrow |5P_{1/2}, F = 2\rangle$ transition, and a second $\sigma^+$ beam coupling the $|5S_{1/2}, F = 2\rangle \rightarrow |5P_{3/2}, F = 2\rangle$ transition) to prepare the atoms in the $|5S_{1/2}, F = 1, m_F = +1\rangle$ state. The second step is a microwave-driven spin flip. A $6.834995\,\text{GHz}$ microwave pulse, with a duration of $15\,\mu\text{s}$, is amplified by a $70\,\text{W}$ amplifier and emitted into the atomic region via a waveguide antenna, driving the atoms into the target state $|5S_{1/2}, F = 2, m_F = 0\rangle$. In the final step, we employ a $10\,\mu\text{s}$ $\pi$ pulse coupling the $|5S_{1/2}, F = 1\rangle \rightarrow |5P_{1/2}, F = 2\rangle$ transition to clear any remaining atoms from the $|5S_{1/2}, F = 1\rangle$ level.

The energy levels applied in the DLCZ scheme are shown in Fig.~2b. We realize the $\Lambda$-type energy level system using the $|g\rangle \equiv |5S_{1/2}, F = 2, m_F = 0\rangle$, $|e\rangle \equiv |5P_{1/2}, F = 2, m_F = 1\rangle$, and $|s\rangle \equiv |5S_{1/2}, F = 2, m_F = 1\rangle$ states of the $795\,\text{nm}$ D1 line. By encoding the spin wave in the magnetic-field-insensitive clock states $|g\rangle$ and $|s\rangle$, we mitigate the impact of environmental electromagnetic noise. This, combined with the collinear optical geometry, ensures a spin-wave coherence time exceeding $500\,\mu\text{s}$.

\subsection*{Section S2. Setup of the Quantum Frequency Conversion}

\begin{figure}
	\includegraphics[width=1\linewidth]{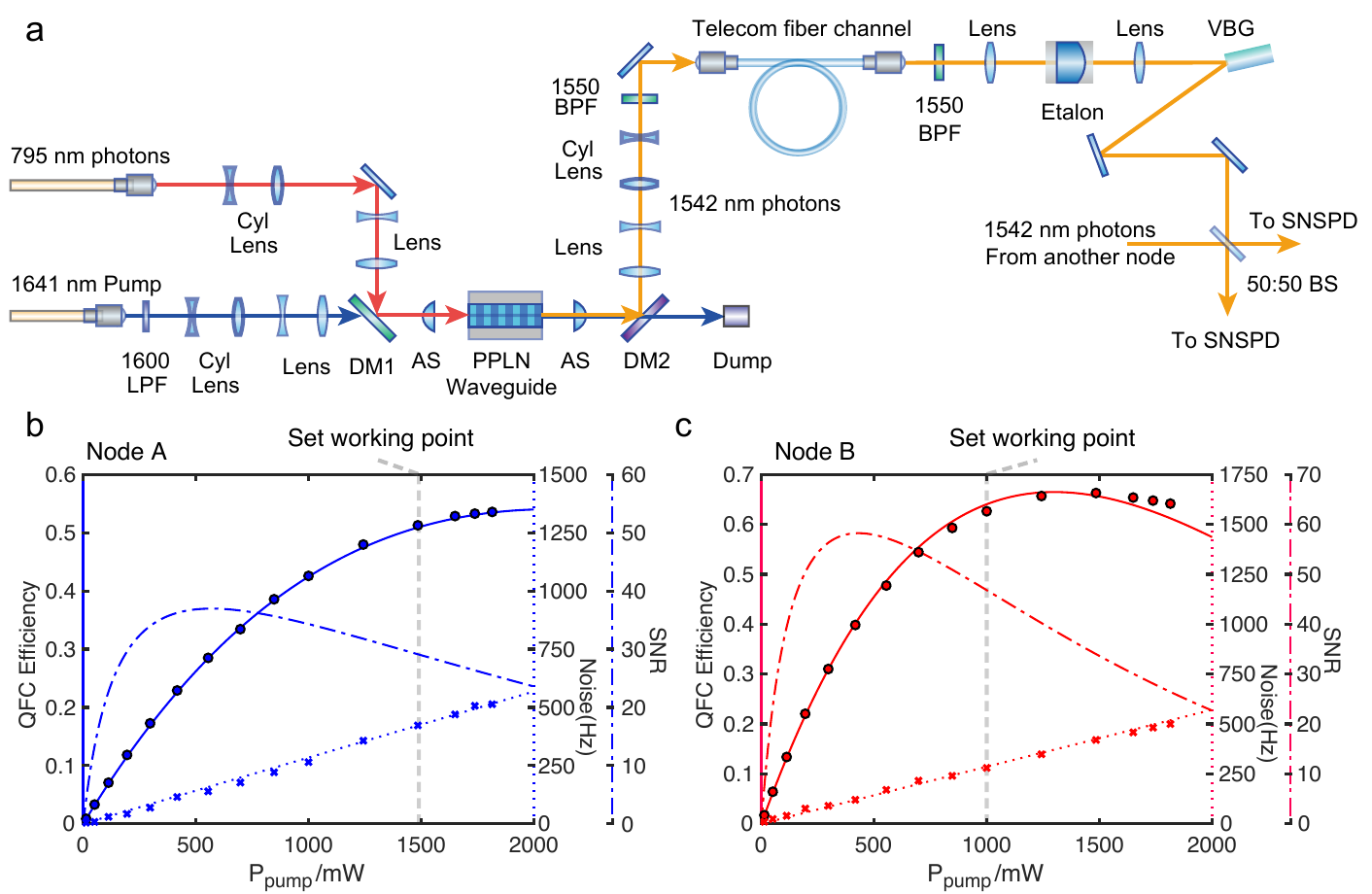}
	\caption{\textbf{Details and performance of the quantum frequency conversion setup.}
		\textbf{a}, Schematic diagram of the quantum frequency conversion setup.
		\textbf{b}, Performance of the QFC module at Node A.
		\textbf{c}, Performance of the QFC module at Node B.
	}
	\label{FigS3}
\end{figure}

In this section, we detail the wavelength conversion module and its performance. The photons generated by the quantum memory array are at $795\,\text{nm}$, a wavelength that cannot be transmitted over long distances through telecom fibers. Therefore, we utilize a difference frequency generation (DFG) process in a periodically poled lithium niobate (PPLN) waveguide. By employing a strong $1641\,\text{nm}$ pump beam, we convert the $795\,\text{nm}$ signal photons into $1542\,\text{nm}$ photons within the telecom C-band. This band experiences minimal transmission loss, thereby facilitating long distance transmission. This specific wavelength combination was selected based on the Raman spectrum of the PPLN waveguide, as it represents the configuration within the target C-band least affected by anti-Stokes scattering noise.

The wavelength conversion module and spectral filtering setup are illustrated in Fig.~S3a. The $795\,\text{nm}$ signal photons and the $1641\,\text{nm}$ pump beam are shaped by a set of cylindrical and spherical lenses, respectively. They are then combined using an $1100\, \text{nm}$ long-pass dichroic mirror and efficiently coupled into the waveguide via an aspheric lens. Through precise beam alignment, the mode-matching efficiency for both beams reaches approximately $90\%$, ensuring high conversion efficiency. The conversion medium is a $2\text{-cm}$-long ridged waveguide, with both end facets custom-coated with a triple-band anti-reflection (AR) coating for $795\,\text{nm}$, $1542\,\text{nm}$, and $1641\,\text{nm}$. The temperature of the waveguide is precisely controlled to maximize conversion efficiency. The waveguide at Node A is maintained at $55.2\,^\circ\text{C}$, while Node B is set to $50.5\,^\circ\text{C}$. The converted $1542\,\text{nm}$ photons are similarly shaped by a set of spherical and cylindrical lenses before being coupled into the $10\text{-km}$ fiber channel to the middle station. To filter out background noise generated during the conversion process and potential stray light in the long fiber, we employ a cascaded spectral filtering module. Both the input and output ends of the long fiber feature a bandpass filter with a central wavelength of $1550\,\text{nm}$, a full width at half maximum (FWHM) of $40\,\text{nm}$, and an extinction ratio exceeding $\text{OD4}$. At the intermediate measuring station, prior to interference, the $1541\,\text{nm}$ photons are further filtered by a plano-convex etalon (free spectral range (FSR) of $13.6\,\text{GHz}$, finesse of $150$, transmission efficiency of $95\%$) and a volume Bragg grating (VBG) with a $10\,\text{GHz}$ FWHM and $98\%$ efficiency.

The specific parameters of the wavelength conversion modules are detailed in Fig.~S3b and S3c. Fig.~S3b presents the module at Node A, which operates under a $1.5\,\text{W}$, $1642\,\text{nm}$ pump beam. It achieves a conversion efficiency of approximately $51\%$ (excluding long fiber coupling efficiency) and a signal-to-noise ratio (SNR) of about $29$. Fig.~S3c presents the module at Node B, operating under a $1\, \text{W}$, $1642\,\text{nm}$ pump beam, achieving a conversion efficiency of approximately $63\%$ and an SNR of roughly $47$. Accounting for the efficiencies of all optical components, fiber coupling, and the detection efficiency of the superconducting nanowire single-photon detectors (SNSPDs), the end-to-end efficiency from the memory array to the intermediate station is $12\%$ for Node A and $15\%$ for Node B.

\subsection*{Section S3. Characterization of phase noise in the long fibers}

\begin{figure}
	\includegraphics[width=1\linewidth]{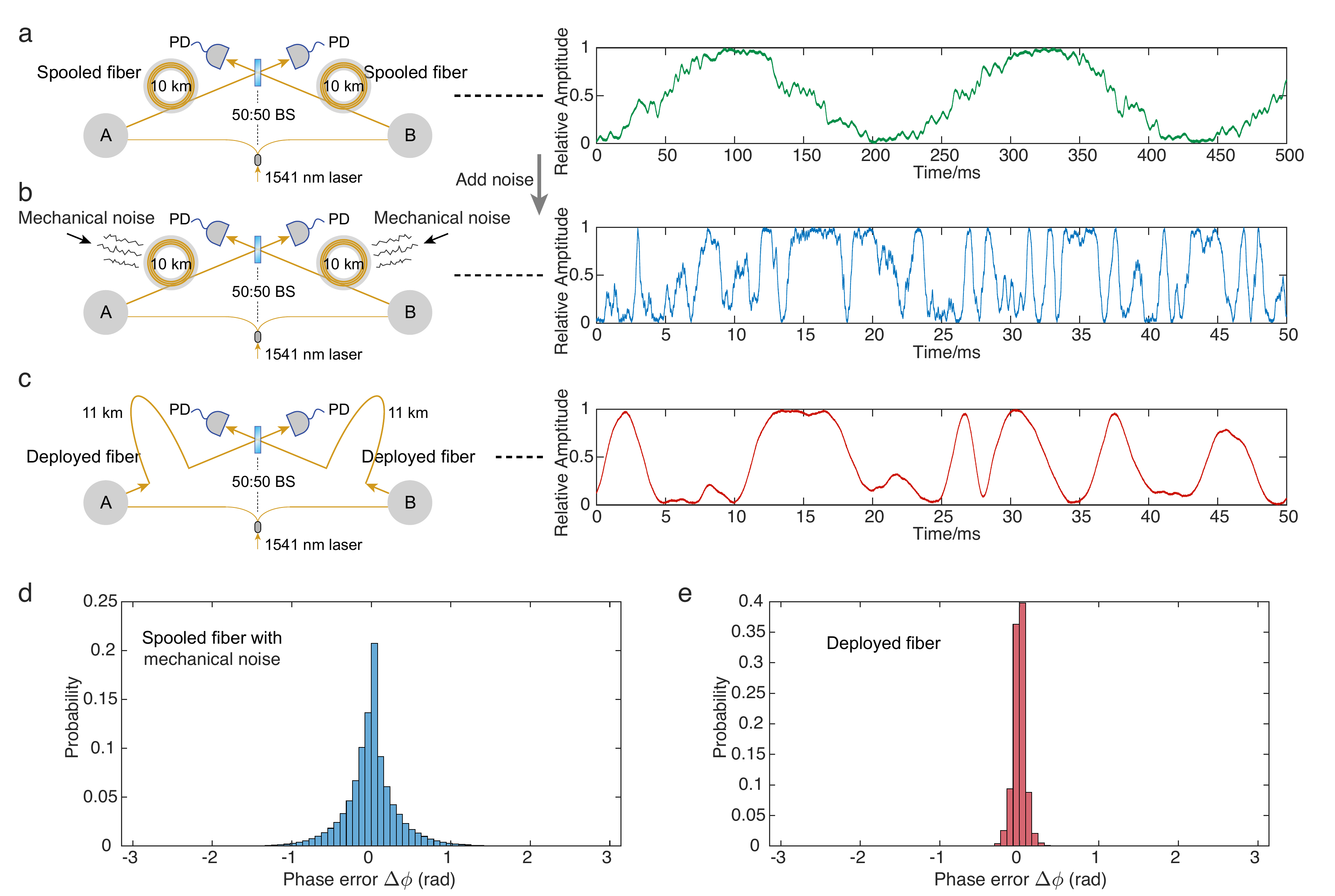}
	\caption{\textbf{Detailed configuration and phase noise analysis of the long optical fiber.}
		\textbf{a}, Interferometric measurement results for the 10-km spooled fiber.
		\textbf{b}, Interferometric measurement results for the spooled fiber with additional applied mechanical noise.
		\textbf{c}, Interferometric measurement results for the 11-km deployed fiber.
		\textbf{d}, Phase error distribution of the spooled fiber with additional mechanical noise used in the experiment.
		\textbf{e}, Phase error distribution when the protocol is applied to the deployed fiber.
	}
	\label{FigS4}
\end{figure}

In this section, we outline the details of the $10\text{-km}$ fiber link and analyze the error induced by the fiber noise. To achieve robust and efficient entanglement generation, we continuously and rapidly excite the memory arrays to sequentially generate two pairs of effective maximally entangled (EME) states at times $t_1$ and $t_2$. Subsequently, these EME states are fused on-demand into a polarization maximally entangled (PME) state. When the entanglement generation rate significantly exceeds the rate of optical fiber phase fluctuations, the influence of phase jitter is effectively suppressed. If the fiber phase drifts by $\Delta\phi$ between $t_1$ and $t_2$, this phase error directly degrades the fidelity of the resulting PME state.

Here, we evaluate the relative phase drift of the fiber channel over time by interfering two $1541\ \text{nm}$ classical beams transmitted through the $10\text{-km}$ setup. Combining this with the experimentally measured time interval between the generation of the two EME states, we can determine the expected infidelity induced by phase fluctuations under different fiber conditions.

We first evaluated the performance of two bundles of spooled fiber; the interference is shown in Fig. S4a. Under these conditions, the characteristic time for a $\pi$ phase flip is approximately $100\ \text{ms}$. This time scale is significantly longer than the typical phase fluctuation time of a $10\text{-km}$ deployed field fiber (approximately $0.1\ \text{rad}$ per $122\ \mu\text{s}$ \cite{riedmatten_fiber}). Such slow fluctuations is insufficient to rigorously verify the feasibility of our scheme for practical, real-world quantum networks.

To rigorously evaluate the robustness of our generation scheme, mechanical vibration noise was deliberately introduced to the fiber spool platform. The channel's response is shown in Fig.~S4b. Under applied vibration, the phase drifts by roughly $0.34\,\text{rad}$ per $100\, \mu\text{s}$, and the $\pi$ phase flip time drops to approximately $1\,\text{ms}$.

For comparison, we conducted the same measurements using $11\text{-km}$ field-deployed fibers. The deployed fiber results (Fig.~S4c) indicate a $\pi$ phase flip time of roughly $5\,\text{ms}$ and a drift rate of approximately $0.1\,\text{rad}$ per $100\,\mu\text{s}$, consistent with previous analyses \cite{riedmatten_fiber}. This confirms that the artificial noise introduced to our spooled fiber is significantly higher than the noise profile of realistic $10\text{-km}$ deployed fibers, making it an excellent testbed for verifying the robustness of our scheme.

Furthermore, we calculated the variation in phase error as the time interval increases. Using the experimentally measured distribution of the time difference between EME state generation events, we determined the expected phase error distribution for our scheme (Fig.~4b and 4c in the main text). When paired within a single round of multiplexed excitation, the phase error distributions are shown in Fig. S4d and S4e. For our artificially noisy spooled fibers, the standard deviation of phase error is $0.3\,\text{rad}$, leading to a resulting infidelity of less than $3\%$. We further projected the performance of our scheme on a practical repeater link using the field-deployed fiber (Fig.~S4e). The projected infidelity drops to roughly $0.14\%$. This verifies the feasibility of our robust entanglement generation scheme in practical environments.

\subsection*{Section S4. Phase Calibration in AOD-addressed Memory Array}

\begin{figure}
	\includegraphics[width=1\linewidth]{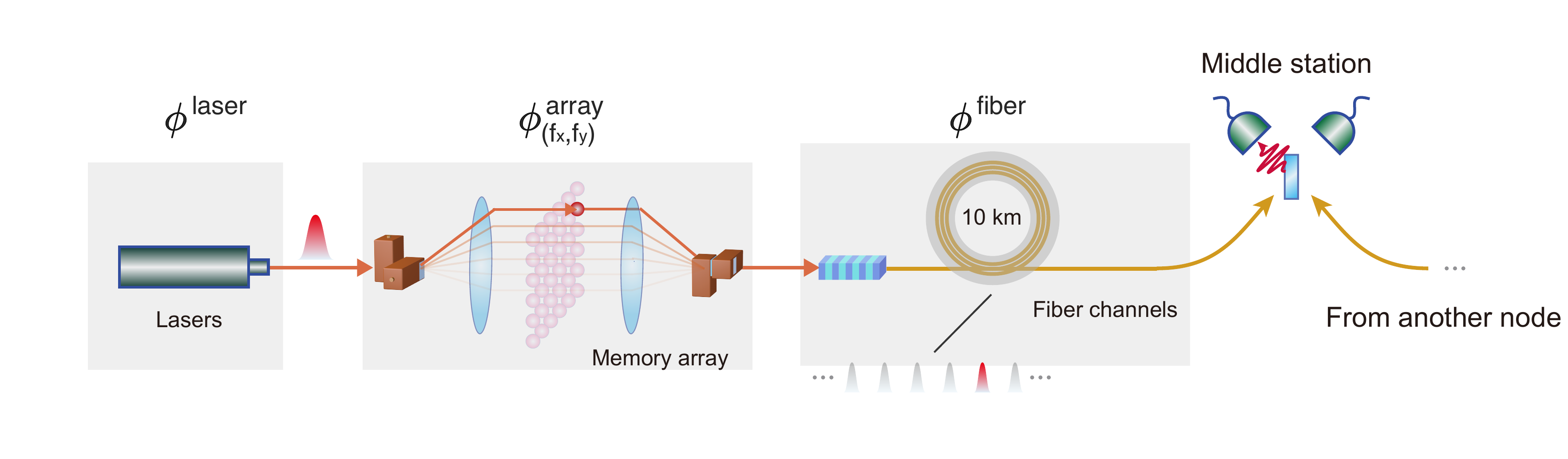}
	\caption{\textbf{Phase components for each node during the single-photon interference process.}
	}	
	
	\label{FigS5}
\end{figure}

This section introduces the phase compensation methodology for different optical paths regarding different cells in the memory array. When utilizing the $(f_x, f_y)$ memory unit across both nodes to generate single-photon entanglement, the heralded EME state is given by:

$$|EME_{(f_x,f_y)}\rangle = \frac{1}{\sqrt{2}}(|10\rangle e^{i\phi_{(f_x,f_y)}^{\text{nodeA}}} + |01\rangle e^{i\phi_{(f_x,f_y)}^{\text{nodeB}}})$$
For each node, the total phase $\phi_{(f_x,f_y)}^{\text{node}} = \phi^{\text{fiber}} + \phi^{\text{laser}} + \phi^{\text{array}}_{(f_x,f_y)}$ comprises three components: the fiber phase $\phi^{\text{fiber}}$, the laser phase $\phi^{\text{laser}}$, and the array-dependent phase $\phi^{\text{array}}_{(f_x,f_y)}$, as illustrated in Fig.~S5. Because all memory units share the same optical fiber and excitation laser, the first two terms are strictly functions of time and are independent of the spatial memory unit. As long as the entanglement generation time is shorter than their respective coherence times, their influence acts as a global phase that is eliminated during the PME state fusion process. (The impact of the fiber phase was analyzed in Section S3, and the laser phase is suppressed by locking the laser to a reference cavity, yielding a sub-kHz linewidth). However, the third term, $\phi^{\text{array}}_{(f_x,f_y)}$, is spatially dependent and varies for each memory unit. It must be carefully calibrated to ensure uniformity across the entire array; otherwise, a phase-stable PME state cannot be generated.

The memory array phase $\phi^{\text{array}}_{(f_x,f_y)}$ encompasses the phase introduced by the radio frequency (RF) signals driving the AODs, alongside the optical path phase unique to each unit. In our setup, spatial addressing during the write and read processes is accomplished by a set of crossed AODs utilizing the +1 diffraction order and a corresponding set utilizing the -1 diffraction order. The $x$ and $y$ dimensions of these AOD sets are driven by synchronized RF signals $f_{x/y}$. Consequently, the total phase they introduce does not evolve dynamically; nevertheless, they introduce a relative phase dependent on the two driving signals, denoted as $\phi^{w/r, x/y}_{f_x, f_y}$. Here, the superscripts $w/r$ indicate the write and read processes, while $x/y$ denote the crossed AOD dimensions. The optical path phase of each unit during these processes is denoted as $\phi^{w/r, \text{path}}_{f_x, f_y}$. This varies spatially and must be compensated for via the phase applied to the AOD RF signals. The total phase introduced by the memory array is:

$$\phi^{\text{array}}_{(f_x,f_y)} = \phi^{w}_{f_x,f_y} + \phi^{r}_{f_x,f_y} = \phi^{w,x}_{f_x,f_y} + \phi^{w,y}_{f_x,f_y} + \phi^{w,\text{path}}_{f_x,f_y} + \phi^{r,x}_{f_x,f_y} + \phi^{r,y}_{f_x,f_y} + \phi^{r,\text{path}}_{f_x,f_y}$$
The calibration procedure is outlined in Fig.~S6a. We designate the ($100\,\text{MHz}$, $100\,\text{MHz}$) unit as the reference, setting all phases associated with the $100\,\text{MHz}$ driving frequency to zero (i.e., $\phi^{w/r,x}_{100,f_y} = 0$ and $\phi^{w/r,y}_{f_x,100} = 0$). For any arbitrary unit $(f_x, f_y)$, we first locate the corresponding reference ($100\,\text{MHz}$, $f_y$) point in its row and adjust $\phi^{w/r,y}_{100,f_y}$ to satisfy $\phi^{\text{array}}_{(100,f_y)} = \phi^{\text{array}}_{(100,100)}$. We then set $\phi^{w/r,y}_{f_x,f_y} = \phi^{w/r,y}_{100,f_y}$ and calibrate the $x$-direction phase $\phi^{w/r,x}_{f_x,f_y}$ such that $\phi^{\text{array}}_{(f_x,f_y)} = \phi^{\text{array}}_{(100,f_y)}$. By systematically applying this across the array, the total phase introduced at every spatial point becomes a constant value, $\phi^{\text{array}}_{(f_x,f_y)} = \phi^{\text{array}}$.

\begin{figure}
	\includegraphics[width=1\linewidth]{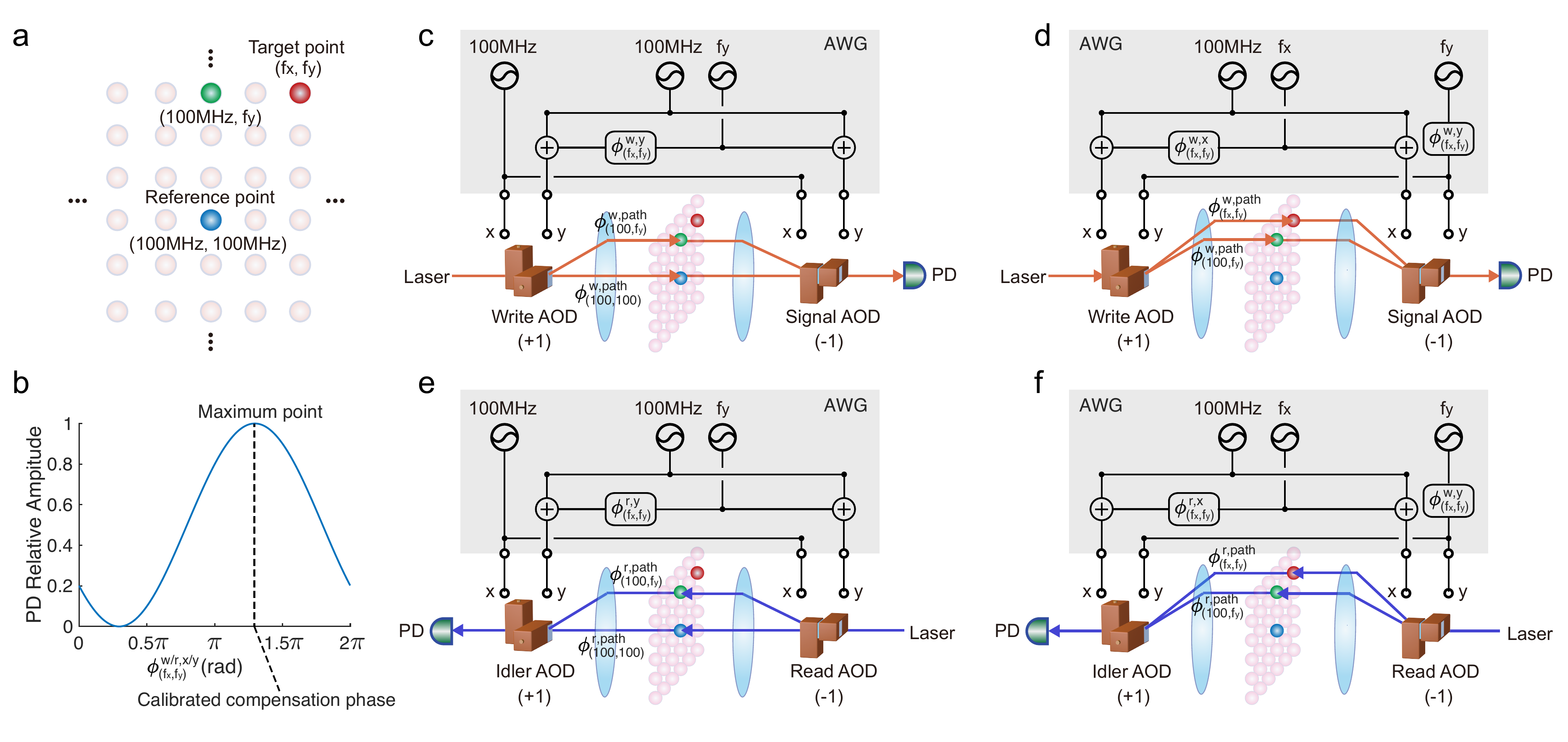}
	\caption{\textbf{Phase calibration within the memory array.}
		\textbf{a}, Configuration for the intermediate reference point.
		\textbf{b}, Phase scanning method and determination of the calibration phase.
		\textbf{c}, Phase calibration in the x-dimension during the write process.
		\textbf{d}, Phase calibration in the y-dimension during the write process.
		\textbf{e}, Phase calibration in the x-dimension during the read process.
		\textbf{f}, Phase calibration in the y-dimension during the read process.
	}
	\label{FigS6}
\end{figure}

We note that there is a degree of freedom in this compensation methodology; successful compensation is achieved as long as the sum of the four AOD phases $\phi^{w/r, x/y}_{f_x, f_y}$ remains constant. By using the $100\,\text{MHz}$ unit of each row as an intermediate reference, $\phi^{w/r,y}_{f_x,f_y}$ remains identical across a given row, while the $x$-direction compensation phase $\phi^{w/r,x}_{f_x,f_y}$ varies within the column. Alternatively, using a $y$-direction reference would invert this dynamic.

We utilize a $795\ \text{nm}$ laser as a phase calibration reference, interfering optical paths traversing two different points on the array. During calibration, we scan the phase between $0$ and $2\pi$. The value maximizing the interference intensity is locked in as the target calibration phase (Fig. S6b), with a scanning resolution of $2^\circ$ ($\approx 0.035\ \text{rad}$).

The write process calibration is detailed in Fig. S6c and S6d. We input the reference beam from the write direction. Using an arbitrary waveform generator (AWG), we apply synchronized $100\,\text{MHz}$ signals to the $x$-dimensions of both the write and signal AODs. In the $y$-dimension, we apply a superimposed signal of $100\,\text{MHz}$ and $f_y$. The $f_y$ signal applied to the +1 order write AOD carries a phase of $\phi^{w,y}_{100,f_y}$ relative to the -1 order signal AOD. This splits the reference laser across the $(100, 100)$ and $(100, f_y)$ paths, which subsequently interfere after the signal AOD. Scanning the phase $\phi^{w,y}_{f_x,f_y}$ to maximize the interference intensity satisfies the condition $\phi^{w,y}_{100,f_y} + \phi^{w,\text{path}}_{100,f_y} = \phi^{w}_{(100,100)}$. We then calibrate the relative phase between the $(100, f_y)$ and $(f_x, f_y)$ paths (Fig. S6d). Scanning $\phi^{w,x}_{f_x,f_y}$ to maximize intensity satisfies $\phi^{w,x}_{f_x,f_y} + \phi^{w,y}_{f_x,f_y} + \phi^{w,\text{path}}_{f_x,f_y} = \phi^{w,y}_{100,f_y} + \phi^{w,\text{path}}_{100,f_y} = \phi^{w}_{(100,100)}$.

The read process calibration strictly mirrors this procedure (Fig. S6e and S6f). Once completed, we apply the total compensation phase $\phi^{\text{compensate}}_{f_x,f_y} = \phi^{w,x}_{f_x,f_y} + \phi^{w,y}_{f_x,f_y} + \phi^{r,x}_{f_x,f_y} + \phi^{r,y}_{f_x,f_y}$. This ensures $\phi^{\text{array}} = \text{const}$, preventing the spatially varying optical paths from imparting any relative phase errors onto the final PME state.

The calibration process only needs to be implemented before the experiment, without the need for calibration in the experiment. The phase remains stable for many days, thanks to the passive stability of the system.

\subsection*{Section S5. Setup During the Entanglement Generation Process}

\begin{figure}
	\includegraphics[width=1\linewidth]{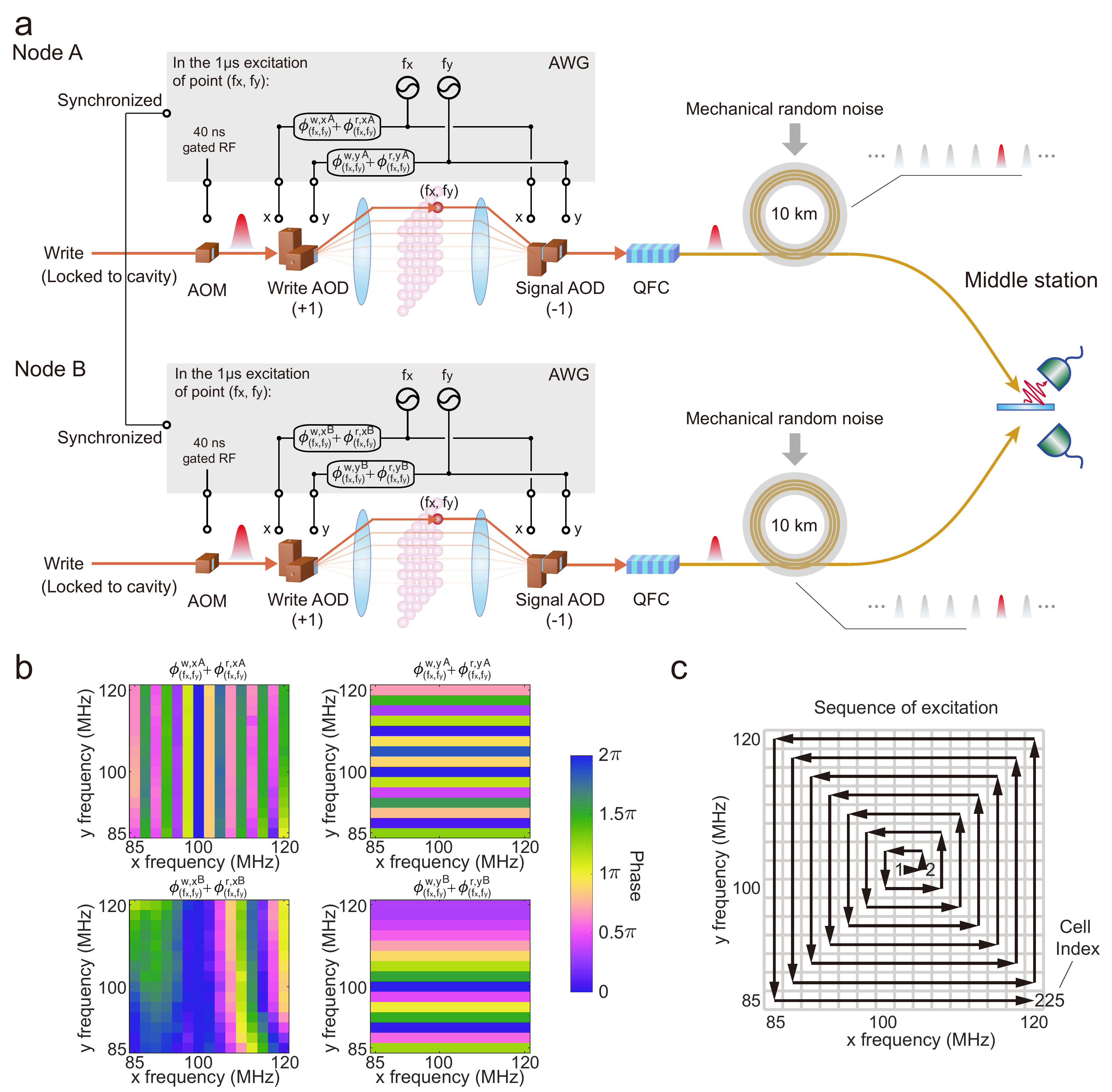}
	\caption{\textbf{Heralded entanglement generation process.}
		\textbf{a}, Configurations for the heralded entanglement generation process.
		\textbf{b}, Compensation phases during the entanglement generation process.
		\textbf{c}, The sequence of scanning the quantum memory array.
	}
	\label{FigS7}
\end{figure}

This section details the control sequence for entanglement generation. During EME state generation, Nodes A and B synchronously sweep their respective memory arrays at a rate of $1\,\mu\text{s}$ per unit, sequentially exciting each cell. The write-out signal photons are converted to $1541\,\text{nm}$ and routed via the noisy $10\text{-km}$ fiber to the intermediate station. Interference measurements at this station herald the generation of entanglement (Fig. S7).

The sweeping and excitation sequence is controlled by an AWG. To excite a specific unit $(f_x, f_y)$, the AWG sends a $40\,\text{ns}$, $160\,\text{MHz}$ RF pulse to the acousto-optic modulator (AOM) controlling the write beam. This slices the cavity-locked write beam into a pulse with an FWHM of approximately $50\,\text{ns}$. Concurrently, the AWG drives the write and signal AODs with $1\,\mu\text{s}$ period at the different $(f_x, f_y)$ frequencies, routing the optical path to the correct cell and subsequently collecting the write-out photons into the channel.

As established in Section S4, stabilizing the total phase requires compensating four distinct terms. We group these into components corresponding to the $x$-dimension ($\phi^{w,x}_{f_x,f_y} + \phi^{r,x}_{f_x,f_y}$) and the $y$-dimension ($\phi^{w,y}_{f_x,f_y} + \phi^{r,y}_{f_x,f_y}$). These are applied as phase differences between the driving frequencies of the +1 order write AOD and the -1 order signal AOD, as presented in Fig.~S7b. This configuration elegantly ensures that spatial phases for both write and read processes are simultaneously compensated during the initial excitation stage. Consequently, during the real-time feedforward in retrieval process, we only need to apply corresponding frequency signals to the read and idler AODs without any dynamic phase adjustments. After the $1\,\mu\text{s}$ excitation window is over, the AWG moves to the next unit, repeating until all the cells of memory array are excited.
Here we note that the time-dependent phases introduced by the driving signals of the acousto-optic deflectors (AODs) in each node's memory array mutually cancel between the write AOD ($+1$ order) and the signal AOD ($-1$ order). Consequently, the memory arrays at different nodes do not strictly need to be excited in the same sequence, and the synchronization between the arbitrary waveform generators (AWGs) controlling the different nodes only needs to be precise to within the photon envelope timescale.

The excitation sequence of the memory array is illustrated in Fig.~S7c. The addressing sequence begins at the central memory cell (located at row 8, column 8, corresponding to the frequency pair ($102.5\,\text{MHz}, 102.5\,\text{MHz}$)) and proceeds outward in a spiral pattern to access the peripheral cells. This particular order is chosen because the central cells exhibit a higher optical depth (OD) and better storage efficiency (as analyzed in Section S1). In our robust PME state generation scheme, the first generated EME state inherently experiences a longer storage time. By utilizing this spiral sequence, we maximize the probability that this first EME state is localized in the highly efficient central cells. This strategy effectively compensates for the extended storage duration, balancing the retrieval efficiencies of the two EME pairs and ultimately yielding a higher overall efficiency.

The generated signal photon pulse train is converted to $1541\,\text{nm}$ and transmitted to the intermediate measurement station. The channel consists of $10\text{-km}$ of spooled fiber subjected to artificial mechanical noise. The signal photons interfere at the intermediate station and are detected by SNSPDs using $60\,\text{ns}$ time windows per unit. Accounting for wavelength conversion and detector dark counts, the overall SNR is $29$ for Node A and $46$ for Node B.

\subsection*{Section S6. Setup for On-Demand Retrieval and PME State Fusion}

\begin{figure}
	\includegraphics[width=1\linewidth]{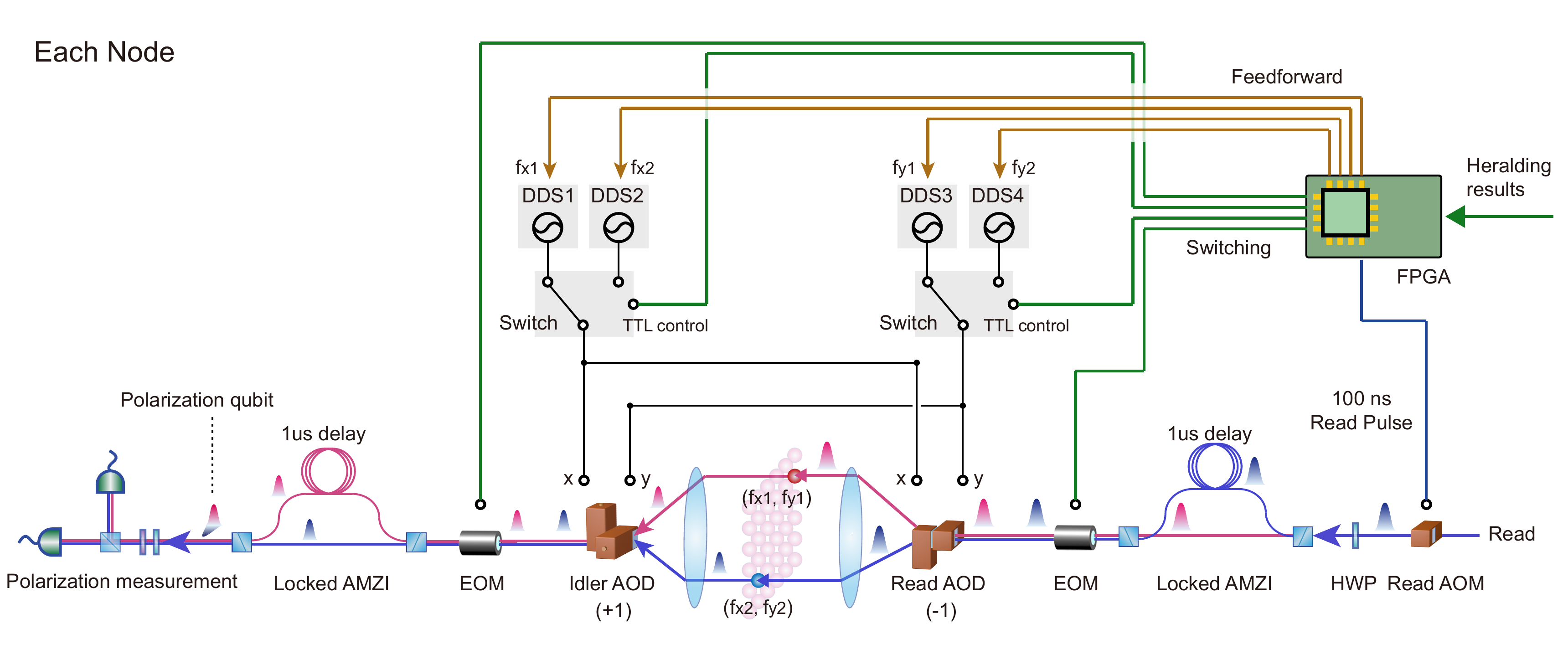}
	\caption{\textbf{Detailed configurations for the read-out and PME state fusion process in each node.}
	}
	\label{FigS8}
\end{figure}

Here, we outline the protocols for real-time feedforward in read-out control and on-demand PME state fusion. A custom-built FPGA at each node manages the dynamic retrieval based on classical heralding signal from the intermediate station. As depicted in Fig.~S8, upon the sequential generation of two EME state pairs, the FPGA identifies the two successfully entangled units via their generation times and computes their AOD addressing frequencies. It encodes these four frequencies (representing the $x$ and $y$ coordinates of the two units) into a $20\text{-bit}$ control signal. This signal is sent at $25\,\text{MHz}$ to four calibrated direct digital synthesizers (DDSs), which rapidly switch to the target frequencies. This feedforward and switching process takes approximately $5\,\mu\text{s}$. Because the path-dependent phase differences were pre-compensated during the excitation sequence, no further phase adjustments are required here. Here we note that the time-dependent phase introduced by each DDS mutually cancels between the read AOD (($-1$
order) and the idler AOD ($+1$ order) and the relative phase has already been compensated during the excitation process, thus no additional synchronization is required among the four DDS modules.

The FPGA then initiates the on-demand retrieval and fusion operations. It carves out a $100\,\text{ns}$ read pulse by controlling the read AOM. This pulse's polarization is rotated to $45^\circ$ via a half-wave plate before entering a time-bin encoding module, converting it into two coherent time-bin pulses separated by $1\,\mu\text{s}$. The module comprises an actively locked asymmetric Mach-Zehnder interferometer (AMZI) and a Pockels cell. The interferometer splits the pulse into two orthogonally polarized pulses separated by $1\,\mu\text{s}$. The FPGA precisely triggers the Pockels cell to half-wave voltage when the delayed pulse passes, matching its polarization to the early pulse. The pulses then enter the array. Through carefully delayed triggering, RF switches alter the read and idler AOD frequencies from $(f_{x1}, f_{y1})$ to $(f_{x2}, f_{y2})$, ensuring the early and late pulses uniquely address and retrieve the respective units one by one.

The retrieved single photons enter a secondary encoding module for fusion into a polarization qubit. Here, the FPGA applies a half-wave voltage to a Pockels cell, routing the early-bin photon into the $1\,\mu\text{s}$ delay arm of an AMZI. The voltage drops before the late-bin photon arrives, allowing it to pass without delay. This precise control ensures the two temporal bins perfectly overlap in time after exiting the AMZI, fusing the two EME states into a robust, polarization-encoded maximally entangled state.

\subsection*{Section S7. Error Budget}

In this section, we analyze the infidelity sources in the finally yield PME state. The infidelity of our system stems primarily from the following sources: readout noise caused by background excitations in the atomic ensemble, noise introduced during wavelength conversion, the occurrence of multiple excitations, amplitude imbalance between terms from different memory cells, and phase errors induced by optical fiber jitter and imperfect phase compensation.

First, we address the impact of background excitations. In our experiment, the excitation probabilities are $\chi \approx 1\%$ (Node A) and $\chi \approx 0.8\%$ (Node B), and an in-sequence cross-correlation of $g_{si}^{\text{array}} \approx 14$. Following wavelength conversion, the signal photons accumulate frequency conversion noise. The modified cross-correlation is evaluated as \cite{riedmatten_QFC}:

$$g_{si} = \frac{g_{si}^{\text{array}} + SNR_{\text{QFC}}^{-1}}{1 + SNR_{\text{QFC}}^{-1}}$$

where $SNR_{\text{QFC}}$ represents the signal-to-noise ratio (SNR) of the quantum frequency conversion module. As detailed in Section S2, the SNR of this conversion process is $29(46)$ at Node A (Node B), leading to a reduction in $g_{si}$ of approximately $0.4$.

During heralded EME generation via single-photon interference, both nodes are excited simultaneously, yet only one spin-wave excitation is heralded. The readout SNR introduced by background noise during a single retrieval event is:

$$SNR_{\text{read}} = \frac{g_{si}-1}{2}$$

This factor of two arises because the statistical weight of a spin wave correlated with the heralded signal photon is $g_{si}-1$, while the weight of an uncorrelated excitation is $1$. Consequently, the probability of successfully reading out a target spin-wave photon is given by:

$$P_{\text{read}} = \frac{SNR_{\text{read}}}{SNR_{\text{read}}+1} = \frac{g_{si}-1}{g_{si}+1}$$

We next incorporate the effects of multiple excitations. In a two-node DLCZ architecture, multiple excitations reduce the EME state proportion by a factor of $3\chi$ and the PME state proportion by $8\chi$ \cite{rmp_gisin}. The fraction of target entangled states successfully extracted during retrieval is therefore:

$$P_{\text{EME}} = P_{\text{read}}(1-3\chi)$$
$$P_{\text{PME}} = P_{\text{read}}^2(1-8\chi)$$

Because the PME state inherently requires the retrieval of two photons, its theoretical fidelity is fundamentally bounded by the square of the readout SNR.

We then evaluate phase errors and amplitude imbalance. The ideal target state is $|\Psi\rangle = \frac{1}{\sqrt{2}}(|HV\rangle + |VH\rangle)$. When a phase error $\Delta\phi$ and an amplitude imbalance factor $\beta$ are introduced, the experimentally fused state takes the form:

$$|\Psi\rangle_{\text{exp}} = \frac{\beta}{\sqrt{1+{\beta}^2}}|HV\rangle + \frac{1}{\sqrt{1+{\beta}^2}} |VH\rangle e^{i\Delta\phi}$$

Here, phase errors arise from long-distance fiber jitter, imperfect phase compensation, jitter in the asymmetric Mach-Zehnder interferometers (AMZIs) utilized during the fusion process, and polarization misalignment in the final measurement basis. Phase jitter in the tens of kilometers of telecom fiber is discussed in Section S3, where an error of approximately $0.3\ \text{rad}$ is introduced by our noise-added spooled fiber. The resolution of the phase compensation process is $2^\circ$ ($\approx 0.035\ \text{rad}$) (detailed in Section S4). Because $4$ independent phases must be calibrated at each node, the accumulated compensation error is approximately $\sqrt{2\times4}\times 0.035\ \text{rad} \approx 0.1\ \text{rad}$. During the retrieval process, $4$ actively locked AMZIs are used; the estimated phase jitter of each AMZI is $0.1\ \text{rad}$, yielding a collective error of $\sqrt{4}\times 0.1\ \text{rad} = 0.2\ \text{rad}$. Finally, polarization misalignment contributes an estimated phase error of up to $5^\circ (\approx 0.085\ \text{rad})$. Assuming these noise sources are statistically independent, the total overall phase error is given by the root-sum-square: $\sqrt{0.3^2+0.1^2+0.2^2+0.085^2}\ \text{rad} \approx 0.38\ \text{rad}$, demonstrating that long-distance fiber jitter is the dominant source of phase uncertainty.

Amplitude imbalance primarily originates from fluctuations in the excitation rates and retrieval efficiencies among different memory cells at each node. These factors produce a final amplitude imbalance estimated to be bounded by $\beta = 1.5$. It is worth noting that neither the global decay of EME retrieval efficiency over storage time nor the baseline efficiency disparity between nodes contributes to this imbalance. This is because each term in the final PME state superposition symmetrically incorporates components from both EME states and both nodes.

The fidelity of this experimental PME state relative to the ideal target is mathematically expressed as $\frac{\beta+\cos(\Delta\phi)}{\sqrt{2}\sqrt{\beta^2+1}}$. Consequently, the final retrieved density matrix, inclusive of noise, is modeled as:

$$\rho_{\text{PME}} = P_{\text{PME}}|\Psi_{\text{exp}}\rangle\langle \Psi_{\text{exp}}| + (1-P_{\text{PME}})\rho_{\text{noise}}$$

Assuming the background and multiple-excitation density matrices $\rho_{\text{noise}}$ represent completely mixed states within the polarization two-qubit Hilbert space (yielding a baseline fidelity of $25\%$), the overall state fidelity evaluates to:

\renewcommand{\arraystretch}{1.5}

$$F_{\text{PME}} = P_{\text{PME}} \times \frac{\beta+\cos(\Delta\phi)}{\sqrt{2}\sqrt{\beta^2+1}} + (1-P_{\text{PME}}) \times 0.25$$

\begin{table}[htbp]
	\centering
	\caption{\textbf{Error budget of the PME state generated.}}
	\begin{tabular}{ll}
		\hline
		\hline
		Sources & Infidelity \\
		\hline
		Read-out noise & 0.175 \\
		Multiple excitation & 0.04 \\
		Telecom fiber jitter & 0.02 \\
		Amplitude imbalance & 0.015 \\
		AMZI Phase fluctuation & 0.01 \\
		QFC background noise & 0.005 \\
		Miscellaneous errors & 0.035 \\
		\hline
		Estimated fidelity & 0.70 \\
		Measured fidelity & 0.70(6)\\
		\hline
		\hline
	\end{tabular}
	\label{tab:S1}
\end{table}

Using this formalism, we calculate the infidelity induced by each factor under our experimental conditions. The error budget is listed in Table S1. The dominant source of infidelity is the read-out noise from background excitations ($\approx 17.5\%$ infidelity), followed by multiple excitations ($\approx 4\%$). The contribution of phase error (mainly telecom fiber jitter) is confined to about $3\%$ by our robust PME state fusion scheme. Furthermore, given that our artificial mechanical noise strictly exceeded realistic field conditions, actual deployment infidelity would fall below $0.5\%$. Amplitude imbalance contributes approximately $1.5\%$ to the overall infidelity. We include other error sources such as unquantifiable systematic drifts, magnetic field noise, and photon wavepacket mismatch at the BS into the miscellaneous errors, which contributes to an infidelity of $3.5\%$. The estimated PME state is $70.0\%$, which is in good agreement with the experimental result $70(6)\%$.

\end{document}